\documentclass{ws-procs9x6}

\usepackage[english]{babel}
\usepackage[latin1]{inputenc}
\usepackage{t1enc}
\usepackage{amsmath}
\usepackage{amssymb}
\usepackage{amsfonts}
\usepackage{graphicx}

\newcommand{\kvec}{\mathbf{k}}

\newcommand{\tauexp}{\left( \frac{3\pi^2}{2}\right)^{2/3}\rho^{5/3}}

\newcommand{\etal}{\emph{et~al}}

\begin{document}

\title{Some challenges for Nuclear Density Functional Theory}

\author{T. Duguet}

\address{National Superconducting Cyclotron Laboratory and Department of Physics
and Astronomy,
             Michigan State University,
             East Lansing, MI 48824,
             USA\\
E-mail: duguet@nscl.msu.edu}

\author{K.~Bennaceur}
\address{Institut de Physique Nucl\'eaire de Lyon,
             CNRS-IN2P3/Universit\'e Claude Bernard Lyon 1,
             43, bd. du 11 novembre 1918,
             F-69622 Villeurbanne Cedex,
             France\\
             CEA/DSM-DIF/ESNT, CEA/Saclay, 91191 Gif-sur-Yvette,
             France\\
E-mail: bennaceur@ipnl.in2p3.fr}

\author{T.~Lesinski and J.~Meyer}
\address{Institut de Physique Nucl\'eaire de Lyon,
             CNRS-IN2P3/Universit\'e Claude Bernard Lyon 1,
             43, bd. du 11 novembre 1918,
             F-69622 Villeurbanne Cedex,
             France\\ E-mail: lesinski@ipnl.in2p3.fr ; jmeyer@ipnl.in2p3.fr}

\begin{abstract}
We discuss some of the challenges that the DFT community faces in
its quest for the truly universal energy density functional
applicable over the entire nuclear chart.
\end{abstract}


\bodymatter

\section{Introduction}

In the study of medium to heavy mass nuclei, nuclear Density
Functional Theory (DFT), based on the self-consistent
Hartree-Fock-Bogoliubov (HFB) method and its extensions, is the
theoretical tool of choice. As new exotic beam facilities are
being built or proposed to be built around the world, DFT is on
the edge of becoming a predictive theory for all nuclei but the
lightest. This is not only true for ground state properties, such
as binding energies, radii or multipoles of the density, but also
for low-energy spectroscopy and decay probabilities.

These advances are possible thanks to the development of better
energy functionals and to the increase of computer resources.
However, the needed accuracy and predictive power for unknown
regions of the nuclear chart still leave a lot of room for
improvements. Thus, DFT faces important challenges in its quest
for the truly universal energy density functional which, in
addition to allowing reliable calculations of extended nucleonic
matter, nuclear masses, beta decay rates or fission lifetimes, is
crucial to the understanding of the astrophysical r-process
relevant to the nucleo-synthesis of heavy elements.

In the present contribution, some of those challenges are
discussed\footnote{This is {\it not} an exhaustive list of
challenges. In particular, it does not cover all methods related
to DFT, i.e. QRPA. Some elements discussed presently reflect the
overall view of the DFT community and constitute the outcome of
the workshop "Towards a universal density functional for nuclei",
September 26-30, 2005, Institute of Nuclear Theory, Seattle, USA.
Of course, some ideas discussed here reflect our personal view on
the matter.}. As a matter of illustration, preliminary results of
an ongoing study addressing the idea of using new {\it ab-initio}
results to adjust the Skyrme energy functional are presented.

\section{Challenges}

The last decade has seen tremendous achievements in developing
microscopic methods to describe medium to heavy nuclei. The most
striking ones have come from nuclear DFT based on the
self-consistent HFB method and its extensions (for a recent
review, see Ref.~\cite{bender03b}). Such methods rely on realistic
effective interactions/functionals (either non-relativistic Skyrme
or Gogny functionals, or relativistic Lagrangians) as the only
(phenomenological) input. Specific correlations beyond the static
mean field are now (almost) routinely incorporated through
restoration of broken symmetries and configuration mixing
calculations (projected-Generator-Coordinate-Method DFT).

Beyond the successes encountered in the development of the method
and in the comparison with experimental data, several formal and
practical challenges still have to be overcome in order to obtain
a unified description of all systems throughout the chart of
nuclei, in particular with the perspective of the construction of
new radioactive beam facilities like RIA.

\subsection{Long term strategy}
\label{strategy}

Energy functional practitioners need a better-defined long term
strategy in the spirit of the "Jacob's ladder" which is referred
to in electronic DFT~\cite{perdew05a}. Indeed, constructing and
adjusting functionals is a tricky game where it has been difficult
to learn from other groups' work. Each group uses its own
construction/fitting strategy, itself evolving in time in a way
which is sometimes not systematic enough to allow for significant
progress.

First, functionals at different levels of the many-body treatment
must be consistent. For example, the HFB and the projected-GCM
functionals must be connected in a clear way, the latter including
the former as a particular case. Thus, projected-GCM DFT should be
motivated from first principle and the HFB level derived from it.
This is still lacking. Second, any improved/complexified
functionals should include all features validated by the
previous/simpler ones. Third, a list of (truly) benchmark problems
should be argued upon and used to make the energy functional
consistent with ab-initio calculations of systems which are
essentially solvable exactly; i.e. Brueckner-Hartree-Fock
(BHF)~\cite{zuo02a} and variational chain summation (VCS)
methods~\cite{akmal98a} in infinite matter, Green Function Monte
Carlo (GFMC)~\cite{pieper01a} method and No Core Shell Model
(NCSM)~\cite{navratil03a} for light nuclei, later followed by
Coupled Cluster (CC) methods~\cite{dean05a} for heavier systems.
An example of that will be touched upon in section~\ref{skyrme}.
Fourth, practitioners need to make progress regarding fitting
algorithms. Systematic errors must be estimated and covariance
analysis of parameters need to be performed to determine their
relevant combinations~\cite{bertsch05a}. Also, one has to assess
the improvements brought about by further ingredients: new theory
input and/or additional experimental data, i.e. superdeformed
states, fission isomers/barriers of (exotic) nuclei. Finally, one
can test new fitting algorithms (re-annealing~\cite{agrawal05a},
genetics) to make sure that the absolute minimum in parameter
space is reached during the fit. It is only following such a
"strict" strategy that significant progress will be made in the
near future.

\subsection{Improved phenomenology}
\label{phenomenology}

Improving single-particle energies coming out of the functional
might be one of the most important subject of focus for the near
future. Indeed, many unsatisfactory single-particle energy
spacings provided at the HFB level spoil features of low-energy
spectroscopy which cannot be corrected for by including further
correlations through beyond mean-field
techniques~\cite{bender04a}. This has not only to do with the
shell evolution towards the drip-lines but also with
single-particle energies in stable systems. In any case, it is
clear that data on particle/hole states in very neutron rich
nuclei will help by providing shell evolution along long isotopic
chains. For instance, information on particle/hole states around
$^{78}Ni$ should be accessible with RIA.

However, there are issues as far as the fitting strategy is
concerned. Indeed, one cannot fit single-particle energies
directly to experiment without including their renormalization
through the coupling to surface vibrations~\cite{bernard80a}. One
would need to adjust functionals through beyond mean-field
calculations of odd nuclei, which is too time consuming to be done
in a systematic way within projected-GCM DFT. One could use other
methods to estimate the effect~\cite{bernard80a}.

One has indications that the previous problem might be partly
related to the omission of the tensor force in DFT. The latter
could explain a lot of closing/opening of sub-shells as one fills
major shells~\cite{otsuka05a}. This is related to the
attraction/repulsion between neutron (proton) $j=l+1/2$ and proton
(neutron) $j =l-1/2/l+1/2$ orbits provided by the tensor
interaction. The $N=32$ sub-shell closure in neutron rich isotopes
($^{54}$Ti) is one possible example. Several groups are now
working on implementing the tensor force in the Gogny and the
Skyrme functionals~\cite{doba06a,lesinski06b,otsuka06a}. It is
important to note that the tensor force has been mostly
disregarded in DFT so far because of a lack of clear-cut
experimental data to adjust it. Again, observation of (relative)
single-particle energies in very exotic nuclei could be of great
help in that respect.

Masses, separation energies, densities, deformation, individual
excitation spectra and collective excitation modes such as
rotation or vibration, depend significantly on the superfluid
nature of nuclei. The role of pairing correlations is particularly
emphasized at the neutron drip-line where the scattering of
virtual pairs into the continuum gives rise to a variety of new
phenomena in ground and excited states of nuclei~\cite{doba3}.

Despite its major role, our knowledge of the pairing force and of
the nature of pairing correlations in nuclei is rather poor. The
structure of the functional as far as gradient versus (isovector)
density dependencies are concerned has to be
clarified~\cite{doba3,duguet2,bulgac3}. Proper renormalisation
techniques of local functionals have been recently proposed but
their content must be further studied~\cite{bulgac1,duguet04a}.
Also, particularly puzzling is the situation regarding
beyond-mean-field effects associated with density, spin, isospin
fluctuations~\cite{shen05a,barranco}. Finally, the influence of
particle-number projection and pairing vibrations has to be
characterized through systematic calculations.

At this point, all modern phenomenological functionals provide
similar pairing properties in known nuclei, i.e. odd-even mass
staggering (OES) between $^{104}Sn$ and $^{132}Sn$, while their
predictions diverge in neutron rich systems~\cite{duguet06a},
where having masses up to $^{146/150}Sn$ or $^{81}Ni$ with an
accuracy $\delta E = 50$ keV would already allow one to
discriminate between most models. Pairing can be probed in other
ways and data on high spin states, low-energy vibrations,
two-nucleon transfer cross sections (feasible with beams of
$10^{4} \,$ part/s) in exotic nuclei will be complementary and all
very valuable.

\subsection{Connection to underlying methods}
\label{connection}

The energy functional has to be rooted into a more fundamental
level, eventually connected to QCD. Doing so is not only important
from a heuristic point of view but also from a practical one.
Indeed, the main problem of current phenomenological functionals
is the spreading in their predictions away from known nuclei. At
the same time, phenomenology indicates that gradient and/or
density dependencies of usual Skyrme functionals are too
schematic/limited (see section~\ref{canauxST}). Thus, we need
guidance beyond a fit on {\it existing} data. Hopes are put into
deriving functionals from Effective Field Theory (EFT) which has
been successfully applied to few-nucleon systems and which offers
a comprehensive picture of the relative importance of three-body
and two-body forces~\cite{hammer05a}. This could be done with the
use of the renormalization group method in the spirit of
$V_{lowk}$~\cite{bogner03a} from which a perturbative many-body
problem seems to emerge in infinite matter~\cite{bogner05a}. The
connection with finite nuclei will call for methods such as the
Density Matrix Expansion~\cite{negele72a}.

\subsection{Grounding beyond-mean-field DFT}
\label{beyondDFT}

Ultimately, the functional has to incorporate, in a microscopic
manner, long range correlations associated with large amplitude
vibrational motion and with the restoration of broken symmetries.
This is of particular importance to finite self-bound systems. The
projected-GCM extension of DFT has proven to be one of the most
promising ways of doing so. For the future, the first challenge
will be to treat more collective degrees of freedom within the
GCM, such as the pairing field or the axial mass octupole moment.
So far, only the axial quadrupole moment has been considered in
systematic GCM calculations~\cite{bender03b}. Second,
configuration mixing and symmetry restoration will have to be
properly formulated, from first principle, within the context of
DFT. While one can use guidance from the case of the average value
of a Hamiltonian in a projected state, DFT presents key
differences which have to be considered seriously. Because of the
lack of theoretical foundation, one has relied on ad-hoc
prescriptions so far to extend the HFB functional to the
projected-GCM level, in particular as far as density dependencies
in the functional are concerned~\cite{duguet03b}. Even worse,
particle-number projected DFT is ill-defined as was recently
realized~\cite{doba05a} and fixed~\cite{bender06a}.

\section{The Skyrme energy functional\:: use of ab-initio inputs}
\label{skyrme}

In relation to section~\ref{strategy}, we now briefly present a
recent attempt to refine the $SLyx$ family of Skyrme
functionals~\cite{chabanat} by using additional ab-initio inputs
from calculations of infinite nuclear matter. Our goal here is to
highlight the general lessons one can learn from such an attempt.

\subsection{Constraining the isovector effective mass $m^{\ast}_{v}$}
\label{effectivemass}

Our original goal was to study the effect of the neutron-proton
effective mass splitting $m^{\ast}_{n}-m^{\ast}_{p}$ with isospin
asymmetry $I= (\rho_{n} -\rho_{p})/\rho$ on the properties of the
Skyrme functional. It is of importance since the effective mass
controls the density of single-particle energies at the Fermi
surface, and hence influences masses and shell correction in
superheavy nuclei as well as how static and dynamic correlations
develop.

While the isoscalar effective mass $m^{\ast}_{s} \, (\approx 0.8)$
is well determined by the ISGQR~\cite{liu76a,reinhard99a},
constraints on the isovector one $m^{\ast}_{v} \, (\approx
0.7-0.9)$ via the IVGDR are not decisive enough. On the other
hand, recent BHF calculations of asymmetric infinite nuclear
matter, with/without $3N$ force, predict $\Delta m^{\ast}_{n-p}
\approx 0.22 \,$ for $I=1$~\cite{zuo01a} ($\, \, \Delta
m^{\ast}_{n-p} \approx 0.13 \,$ for Dirac
BHF~\cite{ma04a,dalen05a}).

Although the amplitude of the splitting cannot be considered as a
ab-initio {\it benchmark}, its sign is more definite. Thus, we
decided to construct Skyrme functionals with different values of
$\Delta m^{\ast}_{n-p}$ with the idea of improving their isovector
properties by using predictions from ab-initio calculations. From
that point of view, it is interesting to note that the $SLyx$
parameterizations were fitted to the Equation of State (EOS) of
Pure Neutron Matter (PNM) with the idea of improving isospin
properties of the functionals. One consequence was to generate
functionals with $\Delta m^{\ast}_{n-p}<0$, in opposition to
ab-initio predictions. On the other hand, older functionals such
as $SIII$~\cite{beiner75a} and $SkM^\ast$~\cite{bartel82a}, which
were not fitted to PNM, had $\Delta m^{\ast}_{n-p}>0$. The same
exact situation happens for the Gogny force~\cite{chappert06a}.
Thus, improving global isovector properties (EOS) seems to
deteriorate state-dependent ones ($m^{\ast}_{v}$) with currently
used functionals.

\begin{figure}[htbp]
\centering
    \includegraphics[width=.49\columnwidth]{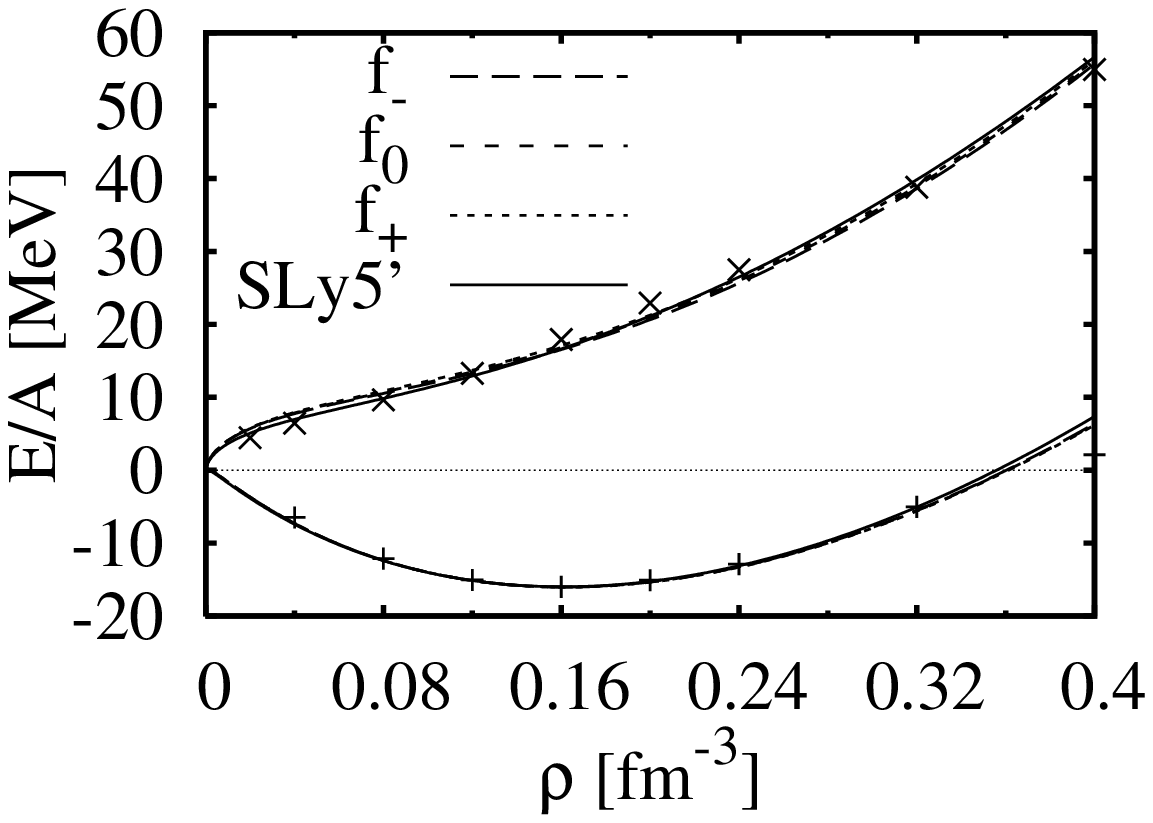}
    \includegraphics[width=.49\columnwidth]{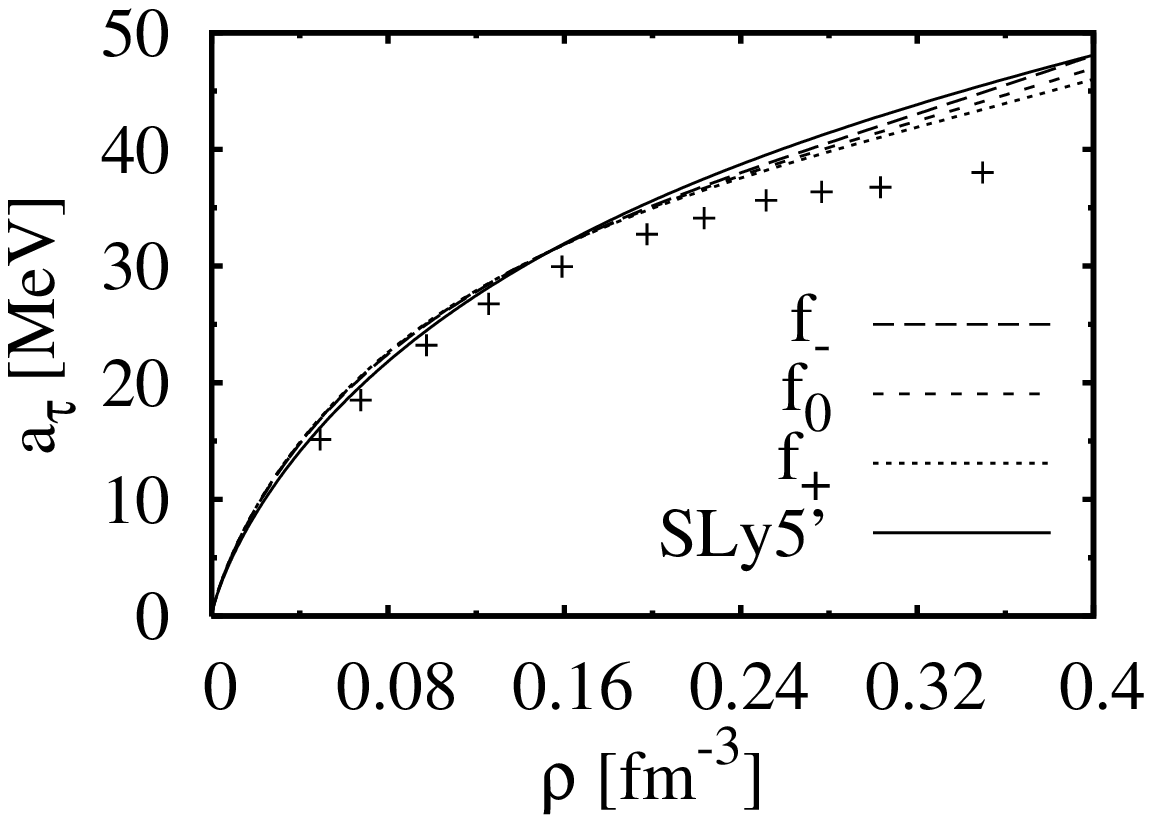}
    \caption{Left\:: SNM and PNM OES as given by
    functionals discussed here (see text), compared with VCS
    results by Akmal~\etal \cite{akmal98a}. Right\:: isospin symmetry energy vs.~density for the
    same set of functionals, compared to variational calculations by Lagaris and
    Pandharipande~\cite{lagaris81a}.}
    \label{fig:eos}
\end{figure}

In any case, we tried to adjust three parameterizations ($f_{-}$,
$f_{0}$, $f_{+}$) using the same fitting protocol as $SLy5$ but
with different values of $m^{\ast}_{v}$. \footnote{The only
difference in the protocol is that static spin-isospin
instabilities below $\rho = 2 \rho_{sat}$ are excluded through the
use of Landau parameters. SLy5' is as SLy5 except for the
ab-initio EOS used in the fit; see Fig.~\ref{fig:eos}.}. Their
bulk properties in SNM are summarized in the table below. They are
all identical and very close to SLy5', except for the effective
mass splitting.

{\small
\begin{center}
\begin{tabular}{ccccccc}
\noalign{\smallskip}\hline\noalign{\smallskip}
 & $\rho_{sat}$ & $E/A_{sat}$ & $K_{\infty}$ & $a_{I}$ & $m^{\ast}_{s}$ & $\Delta m^{\ast}_{n-p}$ \\
\noalign{\smallskip}\hline\noalign{\smallskip}
SkM$^{\ast}$ & 0.160 & -15.770 & 217 & 30 & 0.79 & 0.356 \\
SkP & 0.162 & -15.948 & 201 & 30 & 1.00 & 0.399 \\
\noalign{\smallskip}\hline\noalign{\smallskip}
SLy5' & 0.161 & -15.987 & 230 & 32 & 0.70 & -0.182 \\
\noalign{\smallskip}\hline\noalign{\smallskip}
$f_{-}$    & 0.162 & -16.029 & 230 & 32 & 0.70 & -0.284  \\
$f_{0}$    & 0.162 & -16.035 & 230 & 32 & 0.70 &  0.001  \\
$f_{+}$    & 0.162 & -16.036 & 230 & 32 & 0.70 &  0.170  \\
\noalign{\smallskip}\hline
\end{tabular}
\end{center}
}

First, a few lessons were learnt. Raising $m^{\ast}_{s}/m$ to 0.8
is difficult because of spin-isospin instabilities. With
$m^{\ast}_{s}/m=0.7$, it is impossible to reproduce the high
density part of PNM EOS and fix $\Delta m^{\ast}_{n-p}>0$ at the
same time\footnote{For all formula of relevant quantities, we
refer the reader to~\cite{lesinski06a}.}. It explains why the
$SLyx$ forces predict the wrong sign of the effective mass
splitting in neutron rich matter. Using two density-dependent
terms in the functional $\propto \rho^{1/3}_{0} ;
\rho^{2/3}_{0}$~\cite{cochet04a} allowed the construction of
($f_{-}$, $f_{0}$, $f_{+}$).

In addition, $SLy5$-like functionals with $\Delta
m^{\ast}_{n-p}>0$ generate new finite-size isospin instabilities
which become more dangerous as $\Delta m^{\ast}_{n-p}$ becomes
more positive. The instability tends to split neutron and proton
densities by gaining energy through the term $\propto
C_{1}^{\Delta \rho} \left(\vec{\nabla} \rho_{1} \right)^{2}$ in
the functional, where $\rho_{1}$ is the scalar-isovector density.
The table below gives typical values of $C_{1}^{\Delta \rho}$. We
could check that SkP~\cite{doba84a} already presents such an
instability and that, empirically, $C_{1}^{\Delta \rho}$ must be
greater than $\approx -30 MeV$ to have an instability free
functional. Those finite size instabilities need to be quantified
and better controlled in the fit of the functional. For such a
purpose, we have extended the control to finite wave-length
instabilities by calculating the RPA response function in
symmetric nuclear matter (SNM)~\cite{garciarecio92a,lesinski06a}.

{\small
\begin{center}
\begin{tabular}{cccccc}
\noalign{\smallskip}\hline\noalign{\smallskip}
    & $f_{-}$ & SLy5' & $f_{0}$ & $f_{+}$ & SkP \\
\noalign{\smallskip}\hline\noalign{\smallskip}
$\Delta m^{\ast}_{n-p}$   & -0.284 & -0.182 & 0.001 & 0.170 & 0.399 \\
\noalign{\smallskip}\hline\noalign{\smallskip}
$C_{1}^{\Delta \rho}$  & -5.4 & -16.7 & -21.4 & -29.4 & -35.0 \\
\noalign{\smallskip}\hline
\end{tabular}
\end{center}}

The previous discussion shows already the type of
problems/information arising from our attempt to improve on the
fitting protocol of the Lyon functionals by using more ab-initio
benchmark inputs. Now, the left panel of Fig.~\ref{fig:eos} shows
PNM and SNM EOS as obtained for ($f_{-}$, $f_{0}$, $f_{+}$, SLy5')
and as predicted by variational chain summation
methods~\cite{akmal98a}. The asymmetry energy $a_{I}$ as a
function of density is displayed on the right panel. The message
to extract from such a comparison is that the four forces have the
ability to reproduce both microscopic EOS with the same accuracy.
However, is that enough to characterize the spin-isospin content
of the functionals and assess their quality? In order to answer
such a question, we use further ab-initio predictions and look at
the contributions to the SNM EOS per spin-isospin $(S,T)$
channels.

\subsection{Separation of the EOS into $(S,T)$ channels}
\label{canauxST}

In this section, we discuss the contributions to the potential
energy\footnote{The difference between the total energy and the
kinetic energy of the corresponding free Fermi gas.} of SNM from
the four two--body $(S,T)$ channels. We compare our results with
those predicted by BHF calculations using Argonne V18 two-body
interaction and a three-body force constructed from meson exchange
theory~\cite{baldoperso}. By virtue of antisymmetrization, each
spin--isospin channel is linked to a given parity of the spatial
part of the effective force. However, and as motivated by the DME,
the Skyrme functional is to be seen as the average of an effective
vertex over the Fermi sea~\cite{negele72a}. In particular, the
angular dependence of the original nucleon-nucleon force have been
averaged out. Thus, the fact that only $S$ and $P$ waves are
explicitly considered in the Skyrme functional is illusory. In any
case, one has access to the averaged spin-isospin content by
looking at the different $(S,T)$ contributions to the energy.

Using projectors on spin/isospin-singlet and spin/isospin-triplet
states, it is possible to split the potential energy $E_{p}$ into
$(S,T)$ contributions $E^{ST}_{p}$. In SNM, one finds\::
\begin{eqnarray}
\frac{E^{00}_{p}}{A} &=& \frac{3}{160} t_2 (1-x_2) \tauexp, \label{eq:inmdecompstart} \\
\frac{E^{10}_{p}}{A} &=& \frac{3}{16} \sum_{i=0}^{2}
t_{0i}(1+x_{0i}) \rho^{1+i/3}
    + \frac{9}{160} t_1 (1+x_1) \tauexp, \\
\frac{E^{01}_{p}}{A} &=& \frac{3}{16} \sum_{i=0}^{2}
t_{0i}(1-x_{0i}) \rho^{1+i/3}
    + \frac{9}{160} t_1(1-x_1) \tauexp, \\
\frac{E^{11}_{p}}{A} &=&  \frac{27}{160} t_2 (1+x_2) \tauexp.
\label{eq:inmdecompend}
\end{eqnarray}
where $(t_i,x_i)$ are usual coefficients of the Skyrme functional,
whereas $(t_{0i},x_{0i})$ characterize the density-independent
zero-range term and the two density-dependent ones.

\begin{figure}[htbp]
\centering
    \includegraphics[width=.6\columnwidth]{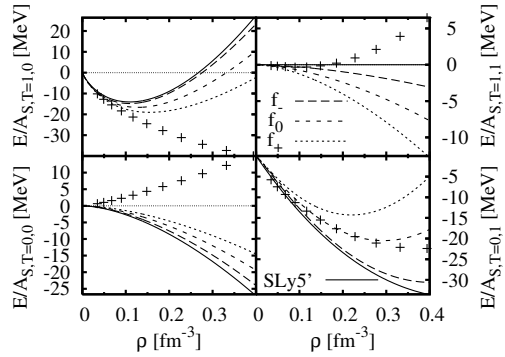}
    \caption{Energy per particle in each $(S,T)$ channel for SNM, as a function of density.
    Crosses refer to the BHF calculations~\cite{baldoperso}.}
    \label{fig:subsp}
\end{figure}

Results are plotted against BHF predictions in
Fig.~\ref{fig:subsp}. First, one can observe that the results are
rather scattered. Second, the main source of binding, from the
$(S,T)=(0,1)$ and $(1,0)$ channels, is not well described and the
saturation mechanism is not captured. It is thus clear that, even
though all four forces reproduce perfectly the PNM and SNM EOS,
they do not have the same spin-isospin content, and that the
latter is in general rather poor. Thus, fitting the global EOS is
an important element but it does not mean that the functional has
good spin-isospin properties. One needs to do more and fitting
ab-initio predictions for the contribution per $(S,T)$ channel
seems to be a good idea. However, one needs to make sure that the
data used are truly benchmarks calculations. This calls for
predictions from other ab-initio methods.

The most obvious discrepancy appears in $(0,0)$ and $(1,1)$
channels where Skyrme and BHF data have opposite signs above
saturation density. The SLy5 parameter set shows a particular
behavior in the $(1,1)$ channel due to the choice of $x_2=-1$ to
prevent ferromagnetic instabilities in PNM. Note that these two
channels are taken care of in the Skyrme functional by the
density-independent $P$-wave term only. The upper--right panel of
Fig.~\ref{fig:subsp} points out the tendency of Skyrme
parameterizations to be attractive in polarized PNM and cause a
collapse of its equation of state at high density. At lower
densities, the BHF data show a distinctive behavior, being
slightly attractive below $\rho_0$ and repulsive above. This
feature cannot be matched by the standard Skyrme functional since
it exhibits in this channel a monotonous behavior with density,
whatever the value of $(t_2,x_2)$.

It is also worth noticing that the failure in the $(1,1)$ channel
becomes more and more prominent as one makes the splitting of
effective masses closer to the ab-initio predictions (force f4).
The effective masses being governed by the momentum dependent
terms of the force, it is not a surprise that the modification of
the former impacts the $(0,0)$ and $(1,1)$ channels. What change
in the coefficients entering
Eqs.~(\ref{eq:inmdecompstart}-\ref{eq:inmdecompend}) stems only
from the variation of $m^\ast_v$ and the associated rearrangement
of parameters in the functional, most notably the $C^{\Delta
\rho}_{0,1}$ coefficients closely related to the surface and
surface--symmetry energies. The quite tight requirements on the
latter imply that the four parameters of the non--local terms in
the standard Skyrme energy functional would be dramatically
overconstrained if we were to add the $(S,T)$--channel
decomposition in the fitting data.

The rather poor properties of the functional in the $(0,0)$ and
$(1,1)$ channels, the degradation of the latter as the effective
mass splitting is improved, the finite-size instability problem,
the idea of using the ab-initio $(S,T)$ contributions in the fit,
call, at least, for a refinement of the odd--$L$ term in the sense
either of a density dependence or a higher--order derivative term.
The latter being prone to numerical instabilities and
interpretation problems, a density--dependent $\kvec'\cdot\kvec$
term remains as one of the next potential enhancements to be
brought to the Skyrme energy functional. Phenomenological
constraints on gradient terms are mainly related to the surface of
nuclei, i.e. low--density regions. One can expect that, to first
order, BHF data in $(S,T)=(1,1)$ channel can be matched with an
extended functional while retaining a good agreement with other
(experimental) data. It is less clear in the $(0,0)$ channel but
further exploration of the extended parameter space may bring
Skyrme and BHF data in better agreement.

\section{Conclusions}
\label{conclusions}

Density Functional Theory faces important challenges in its quest
for the universal energy density functional applicable over the
entire nuclear chart with a true predictive power. In this
contribution, we discuss some of those challenges. Then, we
exemplify one of them by showing the use which can be made of
further constraints from ab-initio calculations and how usual
Skyrme functionals are too schematic/limited as well as over
constrained as soon as one tries to bring more microscopic
character to them.

\section*{Acknowledgments}
The authors thank M. Bender and P.-H. Heenen for many valuable
discussions. Above all, their thoughts go to the memory of Paul
Bonche. This work was supported by the U.S. National Science
 Foundation under Grant No. PHY-0456903.

\bibliographystyle{ws-procs9x6}
\bibliography{riaproc}

\end{document}